\documentclass{aa}  

\usepackage{multirow}
\usepackage{graphicx}
\usepackage{txfonts}
\usepackage{placeins}
\usepackage{siunitx}
\usepackage{hyperref}
\usepackage{xcolor}

\begin{document}

\title{Morphology of Radio Sources in Representation Space}
\author{
Nicolas Baron Perez \corrauth{nicolas.baron.perez@uni-hamburg.de}
\and Marcus Br{\"u}ggen \email{marcus.brueggen@uni-hamburg.de}
\and Luisa Lucie-Smith \email{luisa.lucie-smith@uni-hamburg.de}
}

\institute{Hamburger Sternwarte, Universit{\"a}t Hamburg, Gojenbergsweg 112, 21029 Hamburg, Germany}

\date{Received date / Accepted date }

\abstract
{
Understanding radio source morphologies and their classification remains challenging. We previously developed a deep clustering method based on self-supervised representation learning to classify a subsample of radio sources from the LOFAR Two-meter Sky Survey Data Release 2 (DR2). This yielded a labelled subset used to fine-tune an ensemble of classifiers.
}
{
We aim to identify rare morphological classes in a subsample of LoTSS-DR3, which contains $> 13$ million sources, beyond the 12 classes previously recognised in the DR2 sample. We further aim to characterise the resulting class distribution.
}
{
We applied the classifier ensemble to derive class probabilities and representations for our samples. Among DR3 sources with low class probabilities, we searched for new clusters in representation space. Moreover, we used these clusters as centroids to classify the low-probability subset.
}
{
We found that 88\% of the sources fall into clearly separated clusters in representation space, coinciding with the DR2 clusters. Within the remaining sources, we identified representatives of six additional morphological classes, including rare morphologies like winged and FR-ambiguous sources. The number of sources in the resulting 18 classes exhibits a strongly skewed distribution, with four dominant classes constituting 69\% of the DR3 sample, while rare morphologies account for only a small fraction.
}
{
The identification of additional morphological classes shows the possibility of open-set recognition with representation learning in an unexplored dataset. Unlike anomaly detection, which flags individual sources as uncommon, this approach identifies representatives of novel morphological classes, providing a framework for discovering novel source populations. The highly skewed class distribution poses a fundamental challenge for constructing balanced training datasets and highlights the need for open-set approaches in future observations.
}


\maketitle
\nolinenumbers

\section{Introduction}

The advent of deep, wide-field radio surveys such as the LOFAR Two-Metre Sky Survey \citep[LoTSS;][]{2017A&A...598A.104S}, the Evolutionary Map of the Universe \citep[EMU;][]{2011PASA...28..215N}, and the Karl G. Jansky Very Large Array Sky Survey \citep[VLASS;][]{2020PASP..132c5001L} has dramatically increased the number of radio-loud active galactic nuclei (RLAGN, commonly called radio galaxies) detected with unprecedented angular resolution. In particular, the low-frequency observations provided by LoTSS have revealed previously inaccessible and remarkable features of RLAGN \citep{2022A&A...659A...1S}.

The vast amount of data produced by these surveys, together with the anticipated data from the Square Kilometre Array \citep[SKA;][]{2009IEEEP..97.1482D}, requires extensive testing and development of data analysis methods \citep{2021MNRAS.500.3821B}. Among the various analysis challenges, one key task is the morphological characterisation of RLAGN \citep{2023NewAR..9701685N, 2025A&A...699A.338H, 2025arXiv250919787V}. \\

Since the optical identification of the first radio sources \citep{1949Natur.164..101B}, a wide variety of RLAGN morphologies has been discovered. Several foundational morphological descriptors have played a key role in their classification. One of the most significant is the distinction between centre-bright and edge-bright sources, as introduced by \citep{1974MNRAS.167P..31F, 2019MNRAS.488.2701M} and historically referred to as Fanaroff-Riley (FR) type I and type II, respectively. Another important criterion is the degree of source bending, leading to the classification into wide-angle tail (WAT) and narrow-angle tail (NAT) sources \citep{1980ARA&A..18..165M, 2022AJ....163..280M, 2023Galax..11...67O}. Furthermore, sources exhibiting pronounced lateral extensions or secondary lobes, previously described as S-, Z-, and X-shaped, are now commonly referred to as winged sources \citep{1980ARA&A..18..165M, 2022MNRAS.512.4308B, 2024FrASS..1171101G}.

These classical morphological categories do not strictly partition RLAGN into distinct classes; in most cases, their definitions are based on their appearance rather than on underlying physical processes \citep{1980ARA&A..18..165M}. As the number of detected sources continues to grow, new morphological descriptors have been introduced, often representing combinations or extensions of the initial schemes \citep{2021Galax...9...85R}. A notable example is the class of hybrid sources, which encompasses double-lobed RLAGN that exhibit a centre-bright (FR I-like) morphology in one lobe and an edge-bright (FR II-like) morphology in the other \citep{2000A&A...363..507G, 2020MNRAS.491..803H}.

Recent advances in understanding the life cycle of RLAGN have prompted the introduction of additional morphological categories \citep{2024Galax..12...11M}, including remnant and restarted radio galaxies \citep{2020A&A...638A..34J, 2024Galax..12...11M}.

There is currently no universally accepted classification scheme for RLAGN \citep{2021Galax...9...85R}. As a result, the morphological characterisation of large samples of RLAGN from wide-field radio surveys remains a significant challenge \citep{2025A&A...699A.338H}. Machine learning (ML) offers a promising avenue but is made difficult by specificities of the data and inconsistencies in existing source categorisation approaches \citep{2023NewAR..9701685N}. \\

In the following, we provide an overview of recent advances in ML for RLAGN characterisation; for a broader overview, see e.g., \citet{2025A&A...699A.302B}. Recent developments centre on semi-supervised models, foundation models, vision-language models (VLMs), and alternatives to deep learning (DL) approaches. The semi-supervised models considered here are DL architectures that are first trained on large unlabelled datasets in a self-supervised manner and subsequently trained further (``fine-tuned'') on labelled data \citep{2023arXiv230412210B}. Foundation models represent large semi-supervised models trained on extensive datasets and fine-tuned for multiple learning tasks (``downstream tasks''), allowing them to be applied across a wide range of problems \citep{2021arXiv210807258B}. Vision-language models are multi-modal foundation models that take images and/or text as input \citep{2023arXiv230400685Z}.

Semi-supervised models have been employed for a range of classification tasks under different morphological schemes. \citet{2025MNRAS.tmp.1838B} used the semi-supervised model developed by \citet{2024RASTI...3...19S} to identify a subsample of FR I and FR II sources from the Radio Galaxy Zoo project \citep[RGZ;][]{2025MNRAS.536.3488W}. Their approach relied on a model pre-trained on unlabelled RGZ data and subsequently fine-tuned on labelled FR I/FR II sources \citep{2023RASTI...2..293P}. \citet{2025arXiv251022190H} adopted a similar semi-supervised strategy, pre-training on unlabelled data followed by fine-tuning on a smaller labelled set, to perform binary classification of WAT and NAT sources observed in the VLA Faint Images of the Radio Sky at Twenty-Centimeters survey \citep[FIRST;][]{1995ApJ...450..559B}. \citet{2025A&A...703A.217L} benchmarked vision foundation models, originally trained on natural images, for classification and detection tasks using both optical and radio source images. They evaluated these models either in their unmodified form, with all parameters frozen, or after fine-tuning to astronomical data. Notably, their results show that, while foundation models achieve superior performance in optical source classification, they underperform compared to supervised models in radio source classification. Additionally, \citet{2025arXiv250217207A} explored the explainability of such models by analysing saliency maps generated by various methods in the context of supervised FR I/FR II classification.

VLMs have been systematically investigated for several RLAGN characterisation tasks \citep{2025PASA...42..121R, 2025arXiv250902615D, 2025PASA...42...99G, 2022arXiv221207143C}.

Alternatives to DL models have also been investigated for various RLAGN characterisation tasks. \citet{2025arXiv250518643N} employed COSFIRE (Combination of Shifted Filter Responses) descriptors for the identification of unusual RLAGN. \citet{2026A&C....5401018A} proposed another alternative to DL approaches for the morphological classification of RLAGN (FR I, FR II, compact, bent). \\

Classification models are typically used under the assumption that the datasets to be analysed by the trained model (the inference sets) contain the classes present in the training dataset exclusively. Consequently, all objects in the inference dataset are assigned to one of the considered classes. This standard training paradigm is referred to as closed-set recognition \citep{2021arXiv211006207V}. Therefore, when analysing an entire survey, it is essential that the morphological classes represented in the training data of an ML model reflect the morphological diversity of the detected sources as completely as possible \citep{2025PASA...42...99G}.

Nonetheless, current radio surveys produce continuously growing RLAGN datasets that most likely include new, previously unseen morphologies. This makes it virtually impossible to design a complete classification scheme with the currently available data. Therefore, classifiers should ideally be able to handle morphologies unseen during their training, which is referred to as open-set recognition \citep[OSR;][]{2023arXiv231215571S} in the ML literature. The underlying assumption is that the inference dataset includes classes beyond those present in the training dataset, while the recognition itself can take multiple forms.

OSR and anomaly detection (AD) share the aim of recognising data points in the test or inference datasets beyond the ones known from the training dataset \citep{2021arXiv211014051S} — for an application of AD in radio astronomy, see \citet{2025AJ....169..121L}. However, while AD treats this as a binary task (known vs. unknown), OSR aims at simultaneously classifying known classes and recognising unknown ones \citep{2021arXiv211014051S}. \\

Aiming to obtain a morphologically complete classification scheme for the DR2 sample, we previously developed a deep clustering method that identified 12 morphological classes by combining self-supervised learning with a clustering algorithm, resulting in a labelled subset. An ensemble of classifiers, consisting of the self-supervised encoder and randomly initialised classification heads, was then fine-tuned with this subset, leading to high classification accuracy, reliable uncertainty estimates, and a representation space in which sources of similar morphology cluster together \citep{2025A&A...699A.302B}.

In the present work, we use the classifier ensemble to analyse the DR3 sample, a larger and more diverse dataset than the DR2 sample. We focus on sources assigned low class probabilities, using their learned representations to search for morphological classes beyond the 12 identified in DR2, exploring the capabilities of the model to recognise additional classes without further adaptation (e.g., using an additional OSR-specific loss function). We demonstrate that this method serves as a discovery tool, uncovering representatives of morphologies that were rare or absent in the DR2 sample, without any additional labelling effort. The newly identified clusters are then used as centroids to classify the low-confidence subset, yielding population statistics across all identified classes.

The paper is organised as follows. Sect.~\ref{sect:data} describes the data used in this work. Sect.~\ref{sect:method} briefly recaps the classifier ensemble. Sect.~\ref{sect:additional_morphologies} describes the identification of the additional morphologies in the DR3 sample, characterises them, and discusses the sample's population statistics. Sect.~\ref{sect:conclusions} presents our conclusions.

\section{DR2 and DR3 Samples}
\label{sect:data}

The data originate from LoTSS, a low-frequency survey of the northern sky operating at 120-\SI{168}{MHz} \citep{2017A&A...598A.104S}. We distinguish between two datasets: the ``DR2 sample'', taken from LoTSS-DR2 \citep{2022A&A...659A...1S}, which was used to train the model, and the ``DR3 sample'' drawn from LoTSS-DR3 \citep{2026A&A...707A.198S} and analysed in this work. Both data releases have a nominal angular resolution of \SI{6}{''}, except for LoTSS-DR3 below a declination of $10^\circ$, which has a resolution of \SI{9}{''}. Furthermore, while LoTSS-DR2 covers 27\% of the northern sky, LoTSS-DR3 covers 88\%, corresponding to more than a threefold increase in sky area. Therefore, the DR2 sample is a subset of the DR3 sample.

The two samples were produced in a similar manner. The raw data consist of the Stokes I continuum mosaics, averaged at \SI{144}{MHz}, and the corresponding source catalogues of each data release. The catalogues were filtered to retain only resolved radio sources. The catalogue positions were then used as centres for extracting cutouts from the mosaics. Pixel values below $3 \times \sigma_{\rm cutout}$ were clipped to this threshold, where $\sigma_{\rm cutout}$ denotes the $\sigma$-clipped standard deviation of the cutout. The cutouts were subsequently resized to a common format with a side length of 128 pixels, and min-max normalised. Finally, additional selection criteria (described below for DR3, in \citet{2025A&A...699A.302B} for DR2) were applied to ensure that the selected sources are clearly visible in their preprocessed cutouts.

While for the DR2 sample the radio-optical cross-matched catalogue \citep{2023A&A...678A.151H} was available, for the DR3 sample we used the publicly available PyBDSF \citep{2015ascl.soft02007M} source catalogue\footnote{\url{https://lofar-surveys.org/dr3.html}} (version 1.0) of LoTSS-DR3. This resulted in a slightly different selection function between the two samples, providing a natural test of the model's capabilities to recognise morphological classes beyond those in the training dataset.

The DR3 cutouts differ from the DR2 cutouts in two ways. First, they are centred on the radio mean positions rather than the optical host positions, due to the unavailability of optical identification. Second, they have a side length of $2.5\times\rm{Maj}$, where \rm{Maj} is the major axis of the Gaussian fitted to the radio source, instead of $1.5\times\rm{Maj}$, to accommodate the uncertainties in the size estimates.

The PyBDSF catalogue of LoTSS-DR3 used in this work contains \num{13667877} radio sources, more than three times as many as in LoTSS-DR2 (\num{4116934}). To focus on sources that span more than five beam sizes, we excluded objects with a fitted major axis of $\rm{Maj} < 30^{''}$. This criterion reduced the sample to \num{454086} sources. We then applied the following additional filters:

\begin{itemize}
    \item Sources for which the cutout was not entirely included in the mosaic, contained \texttt{NaN} values, or was empty after clipping were excluded, leaving \num{441015} sources.
    \item We removed all sources with $\rm{Maj} > 60^{''}$, resulting in \num{357528} objects, consistent with the model's training range of $[30^{''}, 60^{''}]$.
    \item To ensure adequate source strength, we required a peak flux density of $F_{\rm peak} > 0.75\,\rm{mJy}/\rm{beam}$; \num{121179} objects satisfied this condition.
    \item To exclude cutouts in which no radio source is visible, we imposed a lower limit of 300 pixels with a pixel value $v_{\rm pix} > 0.1$, retaining \num{110902} objects.
    \item Finally, to remove cutouts dominated by noise, where individual sources cannot be identified, we limited the number of intensity islands to less than 10, resulting in a final sample of \num{108885} sources.
\end{itemize}

The final DR3 sample exceeds the size of the DR2 sample by a factor of approximately 2.6.

\section{Representation learning classifier ensemble}
\label{sect:method}

This section describes the model used in this study to derive image representations, labels, and class probabilities for the preprocessed cutouts.

The model employed in this work was originally developed in \citet{2025A&A...699A.302B} to classify the DR2 sample by characterising its morphological diversity. Its two main objectives were:
\begin{enumerate}
    \item To determine a morphological classification scheme based on the geometrical appearance of the radio sources while minimising the need for supervision. It was achieved through a combination of self-supervised learning, unsupervised clustering, and visual validation. This deep clustering approach yielded a labelled subset.
    \item To produce a lower-dimensional morphological description in the form of representation vectors and to obtain robust class probabilities consistent with the identified classification scheme. For this purpose, the self-supervised encoder was combined with a classification head, and the classifier was fine-tuned multiple times on the labelled subset, resulting in an ensemble of classifiers.
\end{enumerate}

\subsection{Morphological classification scheme}

To project the radio images into a feature space that more compactly encodes their morphological properties, we used the Simple Framework for Contrastive Learning of Visual Representations \citep[SimCLR;][]{2020arXiv200205709C}. This framework takes as input two augmented, i.e., transformed, views of each input image to create positive pairs, while treating all other views in the batch as negatives, outputting representation vectors that are invariant to the used image transformations and whose distances reflect morphological similarity between images.

Within SimCLR, we used a ResNet18 \citep{2016cvpr.confE...1H} as the encoder, producing 512-dimensional representation vectors. To construct augmentations for every input radio image, we used random vertical flip, random rotation, random brightness jitter, and random resized crop, together with the random structural view augmentation introduced in our previous work (pairing morphologically similar sources as positive pairs instead of two views of the same source). The NT-Xent loss function for a pair of projections, $\mathbf{z_i}, \mathbf{z_j}$, is given by:
\begin{equation}
\label{eq:ntxent_loss}
    \mathcal{L}^{\rm NT-Xent}_{\,i,j} = -\log\frac{\exp(\,\textrm{sim}(\bf{z}_i,\bf{z}_j)\,/\,\tau\,)}{\sum_{k=1}^{2N}I_{[k\neq i]}\,\exp(\,\textrm{sim}(\bf{z}_i,\bf{z}_k)\,/\,\tau\,) } \,,
\end{equation}
where $\textrm{sim}(\mathbf{u},\mathbf{v}) = \mathbf{u}^\intercal\mathbf{v}\,/\,||\mathbf{u}||\,||\mathbf{v}||$ represents the cosine similarity, $\tau$ is a temperature parameter, and $I_{[k\neq i]}$ is 1 if $k\neq i$ and 0 otherwise.

We then applied the algorithm Hierarchical Density-Based Spatial Clustering of Applications with Noise \citep[HDBSCAN;][]{2017JOSS....2..205M} to identify overdense regions in the PCA-processed representation space. This yielded a maximum of 95 clusters, which were first filtered using the structural similarity index measure \citep[SSIM;][]{2004ITIP...13..600W} to retain the most homogeneous sets of images. The clusters were then consolidated through visual inspection, either by rejecting spurious clusters or merging similar ones, resulting in a final set of 12 morphological classes.

This procedure resulted in the 12-class scheme listed in the upper part of Table~\ref{tab:class_statistics} (the statistics are discussed in Sect.~\ref{sect:class_statistics}), as well as the labelled subsample used for fine-tuning.

\begin{table}
\centering
\caption{DR3 sample (\num{108885} sources) population statistics.}
\label{tab:class_statistics}
{
\small
\renewcommand{\arraystretch}{1.3}
\begin{tabular}{c|l|r|r}
\hline
Scheme & Name & Count & Percentage \\
\hline
\multirow{12}{*}{\smash{\raisebox{-1.5cm}{\rotatebox{90}{12-class scheme (88\%)}}}}
 & Artifact                   & \num{5407}  & 4.96\% \\
 & Amorphous                 & \num{6812}  & 6.26\% \\
 & Bright core               & \num{2000}  & 1.84\% \\
 & \textcolor{gray}{Head-tail}      & \textcolor{gray}{147} & \textcolor{gray}{0.13\%} \\
 & \textcolor{gray}{Single lobe}    & \textcolor{gray}{27}  & \textcolor{gray}{0.02\%} \\
 & Centre-bright             & \num{5638}  & 5.18\% \\
 & Centrally peaked ellipse  & \num{892}   & 0.82\% \\
 & Symmetric double          & \num{15023} & 13.80\% \\
 & Edge-bright               & \num{39626} & 36.40\% \\
 & Diffuse bent              & \num{13321} & 12.23\% \\
 & Structured bent           & \num{5494}  & 5.05\% \\
 & Circular diffuse          & \num{1365}  & 1.25\% \\
\hline
\multirow{6}{*}{\smash{\raisebox{-0.9cm}{\rotatebox{90}{\shortstack{Additional\\classes (12\%)}}}}}
 & elongated core            & 1155         & 1.06\% \\
 & FR-ambiguous              & \num{2613}  & 2.40\% \\
 & connected edge-bright     & \num{2732}  & 2.51\% \\
 & winged                    & \num{2635}  & 2.42\% \\
 & complex                   & \num{2998}  & 2.75\% \\
 & isolated lobe             & \num{1000}  & 0.92\% \\
\hline
\end{tabular}
}
\tablefoot{A total of 88\% of the sources fall into the 12-class scheme, with the remaining 12\% assigned to the additional classes. Greyed classes are not considered due to DR2/DR3 centring inconsistencies.}
\end{table}

\subsection{Classifier ensemble and representations}
\label{sect:classifier_ensemble}

To build a classifier, the pre-trained encoder was combined with a multilayer perceptron, which mapped the 512-dimensional representations to 12-class probabilities. The classifier was fine-tuned with multiple initial seeds to obtain an ensemble that produces more robust representations and class probabilities. We combined the cross-entropy loss with the Soft Nearest Neighbour Loss \citep[SNNL;][]{pmlr-v2-salakhutdinov07a, 2019arXiv190201889F} into the fine-tuning loss function $\mathcal{L}^{\rm FT} = \alpha\,\mathcal{L}^{\rm Xent} + (1-\alpha)\,\mathcal{L}^{\rm SNN}\,,$ where $\mathcal{L}^{\rm Xent} = -\sum_{c=1}^{n_c} y_c \log(p_c)\,,$ and $\mathcal{L}^{\rm SNN}_i$ is given by:
\begin{equation}
\label{eq:snn_loss}
    \mathcal{L}^{\rm SNN}_i = - \log \frac{ \sum_{j \neq i, \, y_j = y_i,\,\, j=1,...,N} \,\exp(\,\textrm{sim}(\mathbf{z_i},\mathbf{z_j})/\tau^{\rm SNN}\,) }{ \sum_{k \neq i,\,\, k=1,...,N} \,\exp(\,\textrm{sim}(\mathbf{z_i},\mathbf{z_k})/\tau^{\rm SNN}\,) }\,,
\end{equation}
with $y_i$ the class label of the i-th data point and $\tau^{\rm SNN}$ the temperature of SNNL.

With this classifier ensemble, we derived the representations and class probabilities for both samples by averaging the predictions across ensemble members.

\begin{figure*}
\begin{center}
\includegraphics[width=0.95\textwidth]{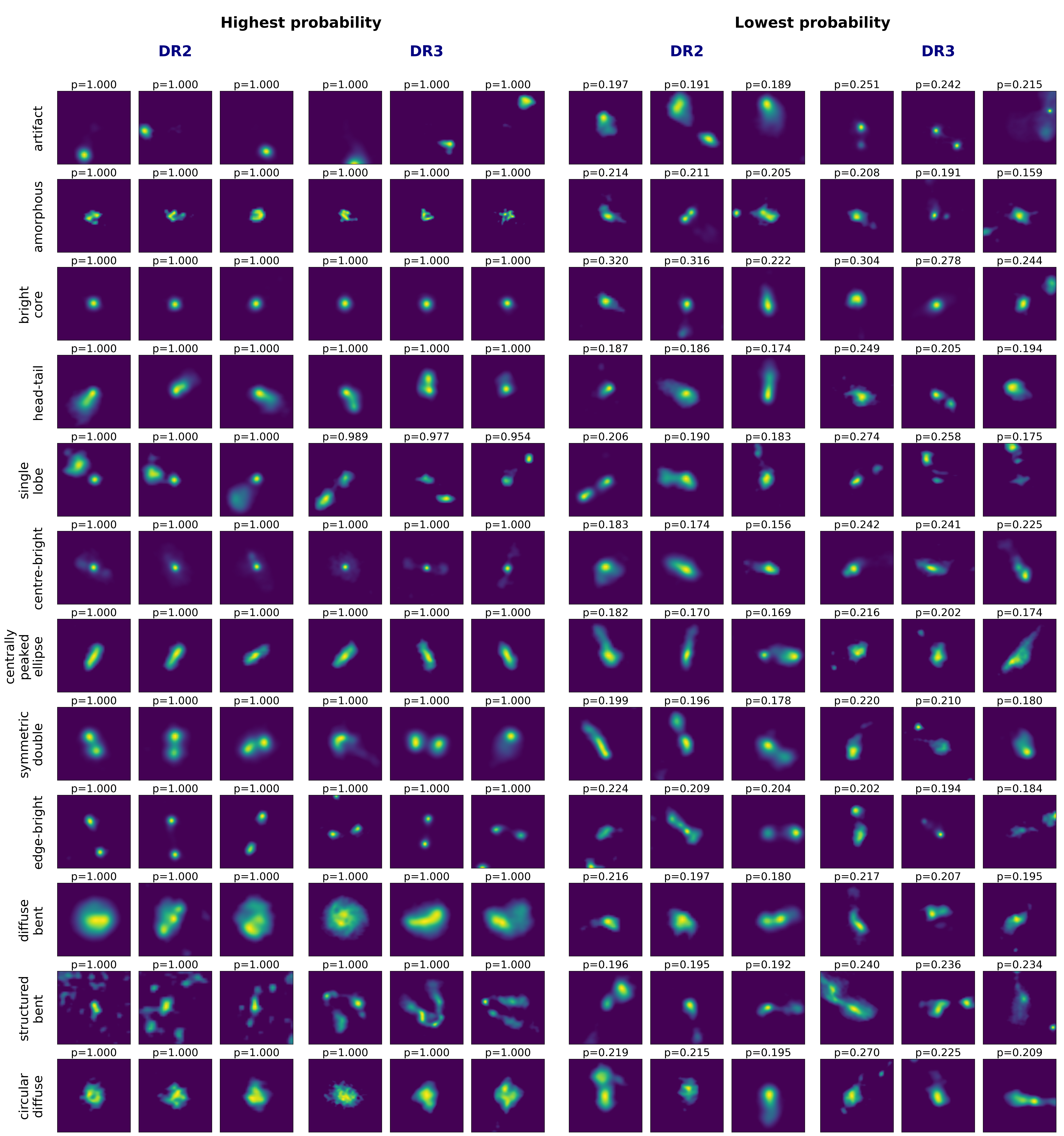}
\end{center}
\caption{Grid of representative images for each class, selected from the DR2 and DR3 samples. Each row corresponds to a specific class. The left half of the columns displays the sources with the highest class probabilities, while the right half shows those with the lowest class probabilities. Within each half, three images from the DR2 sample are presented to the left and three images from the DR3 sample to the right. A few images filled with noise that erroneously passed the data filtering were ignored for this figure.}
\label{fig:class_members}
\end{figure*}

\begin{figure*}
\begin{center}
\includegraphics[width=0.85\textwidth]{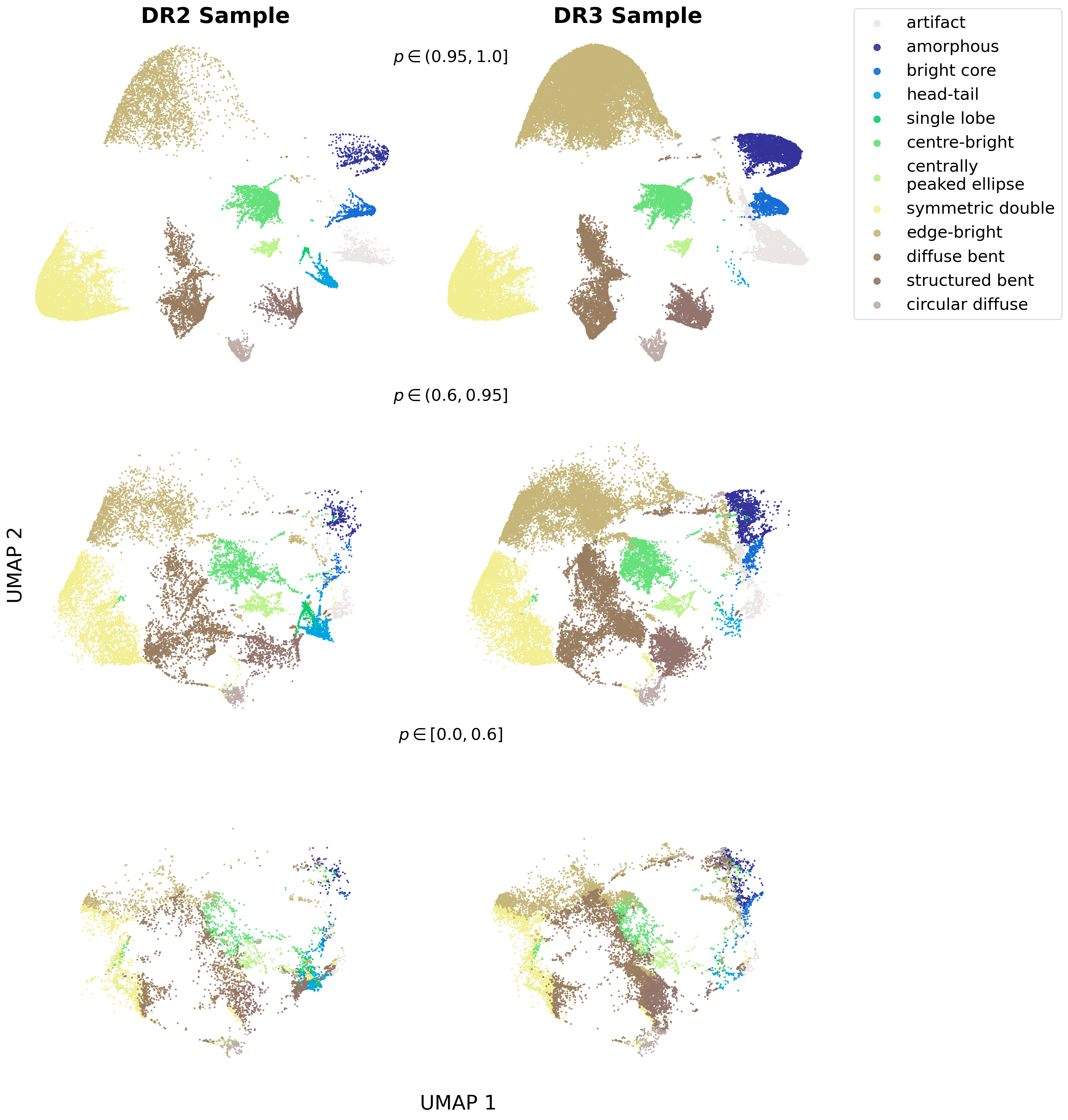}
\end{center}
\caption{Two-dimensional UMAP projections of learned representations, coloured by predicted class, for the DR2 (left column) and DR3 (right column) samples. The data are split into three class probability intervals: $\left[0.0, 0.6\right]$, $\left(0.6, 0.95\right]$, and $\left(0.95, 1.0\right]$ (ordered from bottom to top).}
\label{fig:representations_umap}
\end{figure*}

\section{Additional morphologies and class statistics}
\label{sect:additional_morphologies}

In this section, we classify the DR3 sample, identify six additional morphological classes, characterise their morphologies, and determine the population statistics of the sample.

\subsection{Classification of the DR3 sample}

We predicted the class probabilities for all sources in the DR3 sample using the ensemble of classifiers fine-tuned with the 12-class scheme obtained from the DR2 sample. In Fig.~\ref{fig:class_members}, we show, for both the DR2 and DR3 samples, three representatives from each class with the highest and lowest softmax probabilities to compare the classifications across the samples.

We find that the class probabilities are a good representation of the morphologies for almost all classes. Sources from the DR3 sample with high-probability predictions ($\sim\!100\%$) align closely with the morphologies from the DR2 sample, while low-probability cases ($\lesssim30\%$) often correspond to ambiguous forms. The head-tail and single lobe classes are exceptions: high-probability DR3 sources do not resemble their DR2 counterparts. Most likely, this inconsistency arises because these morphologies are defined relative to the host galaxy position, and centring differs between DR2 and DR3 images. The DR2 cutouts are characterised by a bright dot at the host position and a single diffuse, elongated emission structure corresponding to the overlapping lobes or the single lobe. In contrast, the DR3 cutouts exhibit two or three intensity structures, with one of the outer structures placed at the centre of the image.

We find that a few high-probability symmetric double cutouts exhibit what appears to be a single, diffuse lobe with a potential hotspot. It seems that the model relates these sources to the doubles with a weaker component. Nonetheless, the great majority of the high-probability symmetric double sources have the expected morphology.

We also observed that for the structured bent class, some of the highest probability images are cutouts filled with noise, although the great majority of the class sources from the DR3 sample are consistent with the DR2 sample. These noise-filled images are cutouts from low declination and Galactic regions that passed the data selection filters. They represent examples outside the distribution of the training data, confusing the model.

\subsection{The sample in representation space}

Using the softmax class probabilities, we divided the DR3 sample into low, medium and high probability intervals: $p\in\left[0.0, 0.6\right]$, $p\in\left(0.6, 0.95\right]$, and $p\in\left(0.95, 1.0\right]$. Within these intervals, we find \num{13133}, \num{32974}, and \num{62778} sources, respectively. We show a two-dimensional projection of the PCA-processed representation vectors for each of the probability intervals and for both DR2 and DR3 samples in Fig.~\ref{fig:representations_umap}. Each source in the scatter plots is coloured by its softmax class. The projection was obtained using the manifold learning method Uniform Manifold Approximation and Projection \citep[UMAP;][]{2018arXiv180203426M}.

We find that the distribution of the learned representations aligns closely with the class probabilities. Sources with high-confidence labels form compact clusters with clear separations between classes in the projected space. For sources with intermediate-confidence labels, the clusters begin to merge, but the classes remain largely distinguishable. In contrast, sources with low-confidence labels exhibit significant mixing, resulting in poorly defined clusters. We interpret this as evidence that the 12-class scheme provides an adequate description for approximately 88\% of the DR3 sample (the two highest probability intervals), whereas the remaining 12\% exhibits morphological features that are not well represented by the scheme.

In addition, we find that the distribution of sources in the representation space for the DR2 sample is preserved in the extended DR3 sample, with sources maintaining their clustering according to morphological classes. This demonstrates the transferability of the pre-trained representations across survey data releases. We also observed that the clusters corresponding to individual morphological classes are more fully populated in the DR3 representation space compared to DR2. For instance, some clusters that appear incomplete in the high-confidence subset become more densely populated in DR3 (e.g., the edge-bright class). This indicates the increase in morphological diversity due to the bigger DR3 sample size compared to the DR2 sample. 

\subsection{Identification of additional morphological clusters}

Low-confidence sources occupy poorly separated regions of the representation space and likely include morphologies not well-represented in the twelve training classes, either as transitional types between classes or as genuinely novel morphologies. Identifying these additional morphologies constitutes an OSR scenario: The encoder was trained exclusively on the twelve classes, optimising its representations to discriminate those classes under a closed-set assumption. Consequently, the learned features may not optimally separate additional morphologies absent from the training set. Nevertheless, we hypothesise that sufficiently distinct morphological groups will form identifiable cluster cores.

To find these overdense regions in the representation space of the low-confidence subset, we proceeded as follows: We first applied L2 normalisation to the 512-dimensional representations. Then, we performed principal component analysis (PCA), retaining 95\% of the data variance, which resulted in nine-dimensional projected vectors. These vectors were subsequently clustered using HDBSCAN, with a \texttt{min\_cluster\_size} parameter of 50, a \texttt{min\_samples} parameter of 1, and the Manhattan distance (defined as the sum of the absolute differences of the Cartesian coordinates). This procedure yielded 17 cluster cores, encompassing approximately 32\% of the sources in this subset; the remaining sources were not assigned to any cluster by HDBSCAN.

\begin{figure}
\begin{center}
\includegraphics[width=0.9\columnwidth]{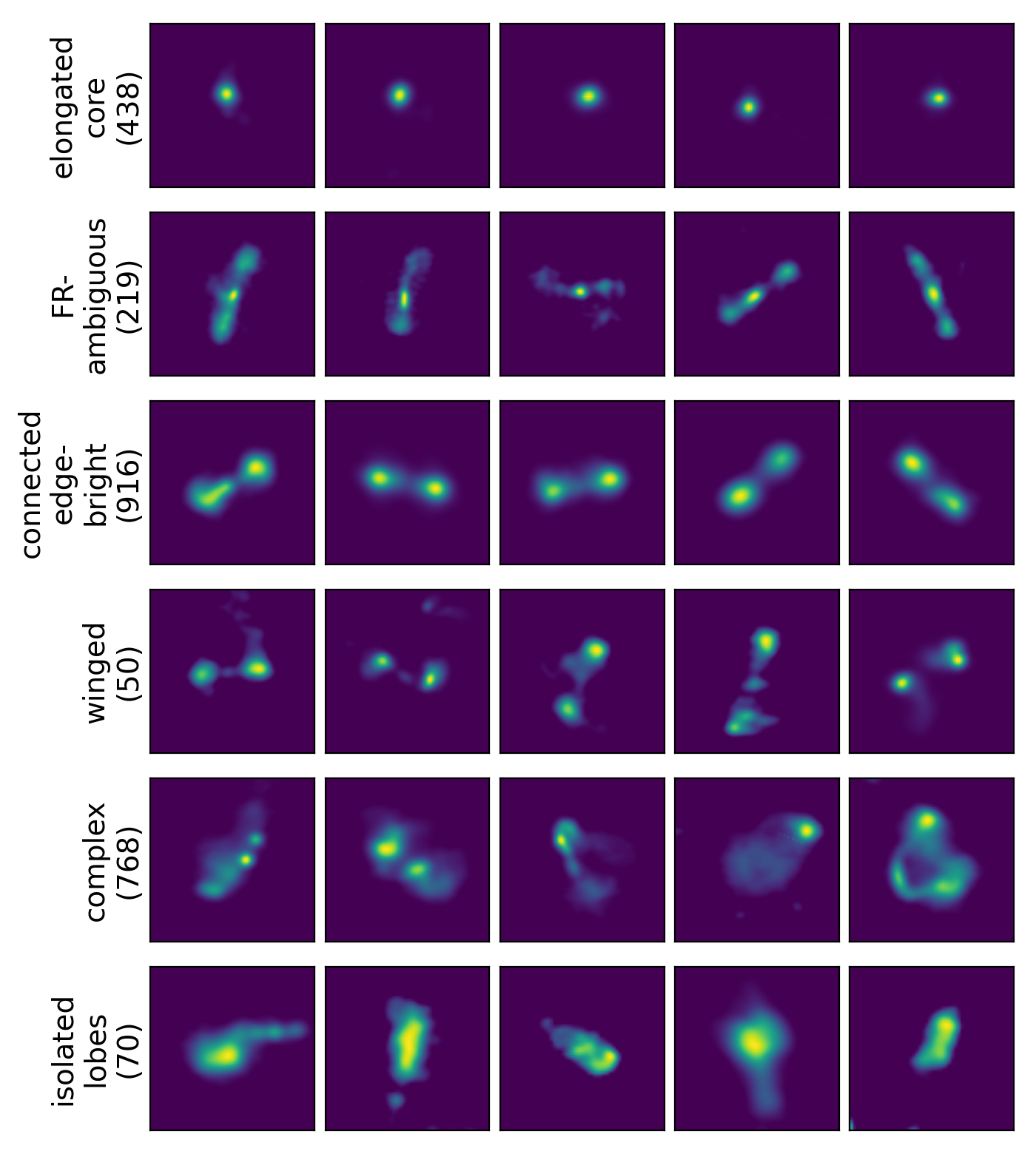}
\end{center}
\caption{A grid of radio source images, with each row illustrating a morphological cluster by showing five representative examples.}
\label{fig:dr3_latent_cluster_examples}
\end{figure}

The visual validation of the clusters led to the exclusion of ten of them. Three clusters contained artefacts and miscentred sources. Three contained sources that shared general features like size, intensity contrast along the source, or number of components but were diverse in radio source-specific properties. The other four clusters included mixed morphologies. Additionally, one pair of clusters with similar morphology was merged. As a result, we identified six additional morphologies to the 12-class scheme. We show five representatives per class with corresponding class names in Fig.~\ref{fig:dr3_latent_cluster_examples}.

The identified clusters are neither complete (they do not contain all representatives) nor entirely pure. However, each cluster contains a sufficient number of representatives to characterise the dominant morphology, as described below and summarised in Table~\ref{tab:cluster_purity}.

\begin{table*}
\centering
\caption{Summary of the properties of the cluster cores exhibiting additional classes. Purity was estimated by visually inspecting a subset or the entire cluster.}
\label{tab:cluster_purity}
{
\small
\renewcommand{\arraystretch}{1.3}
\begin{tabular}{lccl}
\hline
Additional classes & Constituents & Purity & Contaminants \\
\hline
Elongated core & 438 & 0.60 & Bright core, amorphous \\
FR-ambiguous & 219 & 0.85 &  Centre-bright, diffuse bent \\
Connected edge-bright & 916 & 0.63 & Symmetric double, edge-bright \\
Winged & 50 & 0.54 & Edge-bright, diffuse bent \\
Complex & 768 & 0.79 & Common morphologies \\
isolated lobe & 70 & 0.53 & Centrally peaked ellipse \\
\hline
\end{tabular}
}
\end{table*}

The clusters of additional morphologies comprise a total of 2461 sources. Since the LoTSS-DR3 encompasses the observation region covered by LoTSS-DR2, we crossmatched the source catalogues of the DR2 and DR3 samples to estimate the fraction of DR2 sources present in these morphological clusters, thereby showing the importance of the new data release for the identification of the additional morphologies. We used a distance threshold of $\SI{30}{\arcsec}$ for crossmatching and found that 179 sources satisfied this condition. This corresponds to 7.3\% of the sources within the morphological clusters.

To classify the low-confidence subset according to these six morphologies, we employed the nearest-centroid classifier to obtain population statistics for the newly identified classes using the cluster cores identified by HDBSCAN. The algorithm computes the mean position (centroid) of each class in the PCA-processed representation space and classifies unlabelled sources according to the closest centroid. Additionally, we computed the geometric class probabilities by applying a softmax to the scores $s_i = \exp(-\rm{dist}(x, c_i))$, where $\rm{dist}(\cdot)$ is the Manhattan distance between a data point $x$ and the i-th centroid $c_i$.

Using the geometric probabilities, we carried out a qualitative classification validation by visually inspecting the sources with the highest and lowest class probabilities. Sources with high class probability generally matched the intended morphologies, while sources with low class probability showed the expected ambiguity, confirming the method produces sensible assignments despite the limitations.

The class assignments for the remaining 12\% of the DR3 sample in projected representation space are shown in Fig.~\ref{fig:density_latent_classes}. The UMAP projection demonstrates that the nearest-centroid classifier performs well: sources near each cluster core receive the corresponding class label. In the region where no cluster core is located (lower-left), the sources show more heterogeneous labels due to ambiguity.

\subsection{Characterisation of the six classes}
\label{sect:morphological_characterisation}

In this subsection, we characterise each additional class and compare it to the 12-class scheme, both morphologically and in the projected representation space.

\begin{figure*}
\begin{center}
\includegraphics[width=0.74\textwidth]{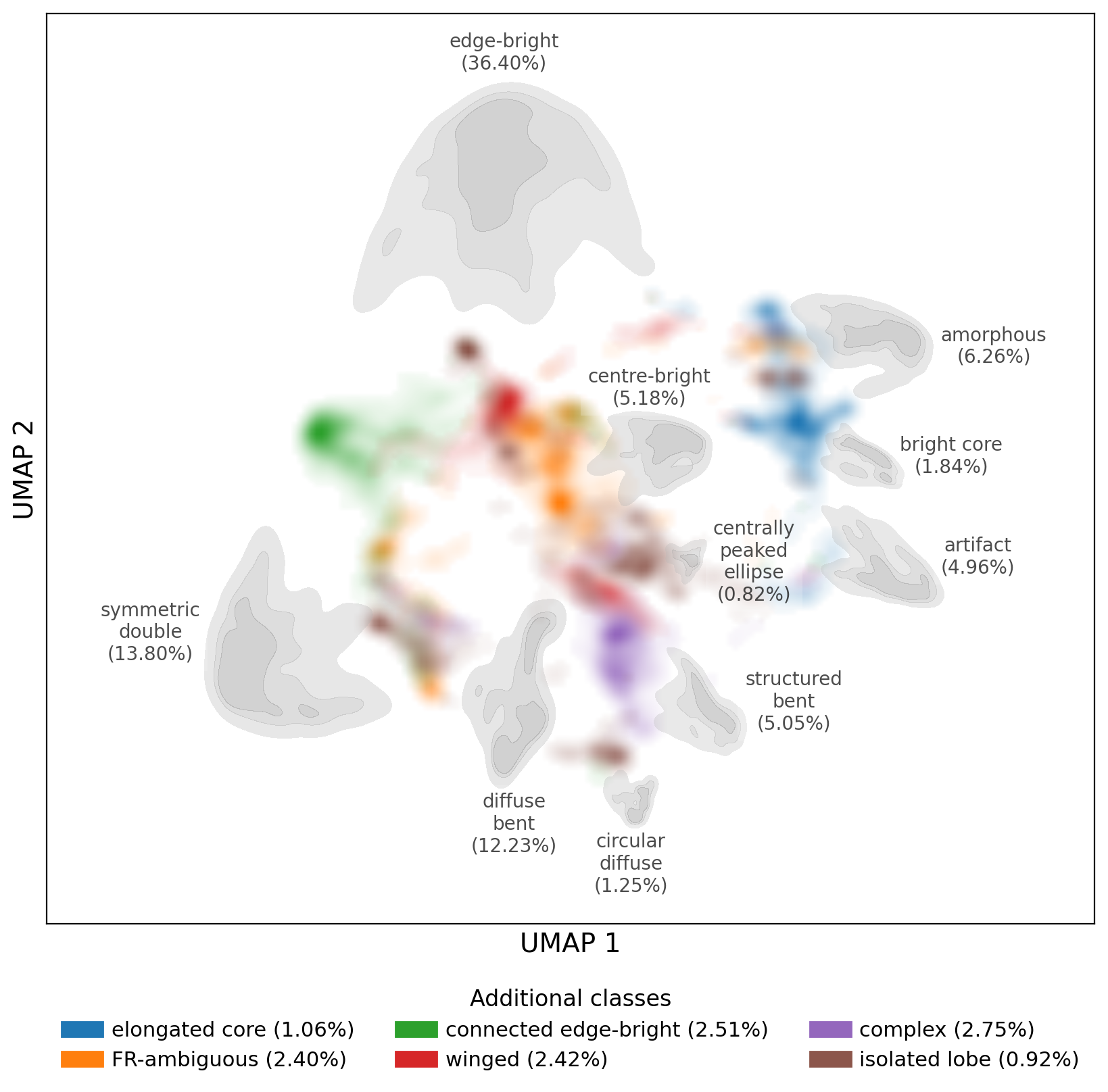}
\end{center}
\caption{Distribution of PCA-processed representation vectors of the DR3 sample in UMAP space. Grey density contours show the high-confidence subset of sources (class probability > 0.95) belonging to the 12 classes (head-tail and single lobe class have been excluded);  although a subset, these contours are representative of the full 88\% of DR3 sources assigned to the 12-class scheme. Gaussian‑smoothed coloured density maps illustrate the remaining 12\% of sources associated with the additional classes. The percentages indicate the fractional contribution of each class to the entire DR3 sample.}
\label{fig:density_latent_classes}
\end{figure*}

In Fig.~\ref{fig:density_latent_classes}, we show the classification of the low-confidence subset according to the newly identified morphological classes as Gaussian-smoothed coloured density maps. For comparison with the 12-class scheme, we also include density contours representing the 88\% of the sample classified under that system. The remaining components of the figure will be discussed later. In the following discussion, we describe the additional classes in the order presented in the figure legend.

\paragraph{Elongated-core sources:} These are sources with bright cores and some diffuse emission that makes them elongated. In the projected representation space, these sources are situated between the bright core, amorphous, and artefact classes, indicating the deviation from spherical symmetry. They span, on average, \SI{37.4}{''} and some of them produce residual sidelobes in the mosaics due to their brightness.

Unlike compact radio sources \citep{2023A&ARv..31....3B}, they show extended emission and are larger than the LOFAR beam. Bright core sources could be centre-bright radio galaxies whose lobes lie along the line-of-sight (potentially BL Lac-type blazars in the unified AGN model). In contrast, elongated-core sources appear to have lobes that are small in angular size and potentially also in physical size.

\paragraph{FR-ambiguous sources:} These sources either comprise three intensity components, the core and two lobes, or these are connected, forming a single component. They have the brightest region at the core but have visible lobes with potential hotspots. At this observation's resolution, it is difficult to determine if the lobes are brightest towards the core, which would make them FR I candidates, or if it is the core, a point source, that appears as an extended feature as it is convolved with the telescope's beam, which would make them FR II candidates with edge-brightened lobes.

In the projected representation space, the cluster is located at the centre, close to the centre-bright and the centrally peaked ellipse classes, extending from the direction of the edge-bright class towards the diffuse bent class. The centre-bright and centrally peaked ellipse sources possess highly dominant cores; the diffuse bent class is characterised by having the core connected with the lobes, and the edge-bright sources have the most prominent lobes.

These sources could be low-luminosity FR II (FR II-lows), which have been found to have a higher prevalence of cores as compact features than FR II-highs \citep{2026MNRAS.549ag894B}.

\paragraph{Connected edge-bright sources:} The sources are characterised by two hotspots at the endpoints of the lobes, which are connected with a dimmer radiation bridge. These hotspots contribute most to the source's brightness. The cluster is positioned in latent space between the symmetric double and the edge-bright class, indicating the intermediate morphological characteristics between the two classes.

The linear shape of these sources is indicative of the absence of interaction with their environment, suggesting that they reside in sparse regions, as is typical for edge-bright sources \citep{2017MNRAS.466.4346M}. While the presence of radiation connecting the hotspots could be indicative of strong, visible backflow in the lobes \citep{2026MNRAS.tmp.1149L}.

\paragraph{Winged sources:} These sources are edge-bright sources with wing-like lobes of diverse shapes. These types of sources are referred to as winged RLAGN \citep{2002MNRAS.330..609D, 2007AJ....133.2097C}, and depending on the shape, they are further characterised as x-shaped \citep{2024FrASS..1171101G}, or s-/z-shaped \citep{2003ApJ...594L.103G}. This cluster is located mainly between the edge-bright and centre-bright classes.

This morphological class is a rare one, with estimates from the finalised FIRST survey ranging from $\sim 2\% - 6\%$ \citep{2019ApJS..245...17Y}. Currently, there is no single model that explains the formation of these sources; the origin is likely a combination of three mechanisms: hydrodynamic backflow in an asymmetric atmosphere, jet reorientation due to spin-flip or precession, and jet intermittency and wobbling from accretion dynamics \citep{2024FrASS..1171101G}.

\paragraph{Complex sources:} These sources exhibit prominent complexity, exhibiting more diffuse emission, multiple bending points along the source, and characteristics exceeding the aforementioned ones. The cluster includes a variety of morphologies: Winged sources, (strongly) bent centre- and edge-bright sources, isolated lobes, W-shaped sources, head-tail sources, and other particular examples. The cluster is found between the diffuse bent and structured bent classes, closer to the latter, as would be expected for sources with highly complex morphologies. The high complexity of these sources is a sign of strong interaction between these sources and their environment, making it likely that most of them lie in merging galaxy clusters.

\paragraph{Isolated lobe:} These sources are predominantly characterised by a narrow, collimated intensity structure (the jet) at one end, transitioning into a more diffuse, elongated emission region (the lobe) that may contain an embedded hotspot. For some of the sources, a compact Gaussian-like intensity peak (the core) close to the jet is visible. In projected representation space, this cluster lies closest to the circular diffuse class and extends from the structured bent to the diffuse bent class. This positioning encompassed both the diffuse nature of the circular diffuse class and the deviation from the spherical symmetry typical of the structured bent class. Checking larger sky regions, we identified that these cutouts contain dominantly isolated lobes of larger radio sources.

These isolated lobes were present as single sources in the catalogue version used in this work and were therefore processed as any other radio source. As isolated lobes, they exhibit a distinctive morphology that differs significantly from sources present in the DR2 training data, providing a compelling example of a class beyond the training dataset.

\subsection{Population statistics of the DR3 sample}
\label{sect:class_statistics}

We provide the class statistics for the DR3 sample in Table~\ref{tab:class_statistics} and in Fig.~\ref{fig:density_latent_classes}. The classification of the high- and intermediate-confidence subsets (88\% of the DR3 sample) is shown in the upper part of Table~\ref{tab:class_statistics} and next to the grey densities of Fig.~\ref{fig:density_latent_classes}, while the population statistics for the low-confidence subset (the remaining 12\% of the sample) are presented in the lower part of Table~\ref{tab:class_statistics} and in the legend of Fig.~\ref{fig:density_latent_classes}.

In agreement with the visual validation, we find that the head-tail and single lobe classes are almost unpopulated, comprising 0.13\% and 0.02\% of the DR3 sample. We therefore discarded these classes from the rest of the analysis.

The edge-bright class is the most populated one with 36.4\% of the sources, followed by the symmetric double class with 13.8\%. We find that the classes overall follow a strongly skewed distribution, with these two classes already containing 50.2\% of the DR3 sample. The third most populated class is the diffuse bent class with 12.2\%, and adding the centre-bright sources, which make up 5.2\% of the sample, these common classes encompass $\sim68\%$ of all sources, leading to low numbers for the remaining classes.

We obtain a ratio of centre-bright to edge-bright sources of approximately 0.14. Combining the centre-bright and centrally peaked ellipse classes into the centre-bright superclass, and the edge-bright and symmetric double classes into the edge-bright superclass, the ratio is 0.12. Other classification studies likewise report a higher number of edge-bright than centre-bright sources, but with larger ratios of 0.65 \citep{2025MNRAS.541.3452C}, 0.48 \citep{2025A&A...699A.338H}, and 0.31 \citep{2025A&A...699A.302B}.

The structured bent sources encompass $\sim5\%$. This leads to a ratio of bent over linear centre and edge-bright sources of 0.42. Other studies lead to ratios of 0.3 \citep{2025A&A...699A.338H}, and 0.4 \citep{2025A&A...699A.302B}. Moreover, we find a significant amount of spurious cutouts of $\sim11\%$ (the artefact and amorphous classes together). The bright core, circular diffuse, and centrally peaked ellipse classes are the least populated classes in the sample.

From the newly identified morphologies, the complex class is the largest one, comprising $2.75\%$ of the DR3 sample, while the isolated lobe class is the smallest class with 0.92\%. The high fraction of complex sources in the low-confidence subset is consistent with the difficulty to characterise these subset with the 12-class scheme.

\section{Conclusions}
\label{sect:conclusions}

We have identified additional morphologies in the DR3 sample and classified it according to a resulting scheme of 18 classes. The distribution of radio source morphologies follows a strongly skewed distribution, with four classes (edge-bright, symmetric double, diffuse bent, and centre-bright) containing 77\% of all sources, while rarer classes such as winged or FR-ambiguous morphologies account for around 2.5\% each. This intrinsic (astrophysical) distribution challenges standard ML classification approaches and requires alternatives such as long-tailed learning \citep{2024arXiv240800483Z}. Although identifying these rare and complex morphologies is challenging, they are expected to shed light on RLAGN evolution and their influence on galaxy formation. This makes ML techniques capable of recognising classes beyond the training set, such as open-set recognition and anomaly detection, crucial for the analysis of future surveys.

In previous work, we developed a deep clustering approach based on self-supervised learning and unsupervised clustering to characterise the morphological diversity of the DR2 sample. This yielded a 12-class scheme and a subset of labelled radio sources. The self-supervised encoder and the labelled sources were then used to fine-tune an ensemble of classifiers tailored to radio surveys. In this work, we applied the ensemble to the DR3 sample to derive class probabilities and representations, and searched for novel morphological clusters among sources with low class probabilities. Without further model adaptation, we identified six additional morphologies after visual validation, demonstrating the capability of the ensemble to recognise classes beyond the training dataset and to identify rare morphologies in evolving datasets. We then classified the low-probability sources according to these six classes and provided class statistics for the DR3 sample.

The application of the 12-class scheme to the DR3 sample showed good generalisation across the data releases despite a 2.6-fold increase in source number (the DR2 sample is included in the DR3 sample), as supported by two comparisons. First, the comparison of sources with the highest class probabilities in both samples (Fig.~\ref{fig:class_members}) showed good agreement for the majority of classes, with two exceptions due to data preprocessing differences. Second, the comparison of the representations of the DR2 and DR3 samples (Fig.~\ref{fig:representations_umap}) showed that the clustering according to morphological classes from DR2 is maintained in the DR3 sample, with additional sources populating the class regions due to the larger sample size, and the classes remaining separable in representation space for 88\% of the DR3 sources. This motivated and validated the search for morphological clusters among the low-class probability sources. 

Conventional classification approaches rely on manually labelled subsets of radio sources, which ensures high-confidence labels. However, these subsets contain a limited number of classes and their selection function is usually unknown, making the closed-set assumption (that the inference dataset contains the classes present in the training dataset exclusively) unlikely to hold when applying the classifier to a survey subset. In contrast, our approach provides a more principled setting for studying morphological classes beyond the training dataset, highlighting the advantage of our classification scheme.

The additional morphologies include genuinely rare morphologies and a cluster containing isolated lobes (see Fig.~\ref{fig:dr3_latent_cluster_examples} for cluster examples). These clusters lie in morphologically meaningful positions relative to the 12-class scheme, as shown in Fig.~\ref{fig:density_latent_classes} and discussed in Sect.~\ref{sect:morphological_characterisation}. They are neither complete nor entirely pure, but each cluster contains sufficient representatives to characterise its dominant morphology (Table~\ref{tab:cluster_purity}). These representatives offer candidates for follow-up studies and enable morphological discovery in growing surveys without additional labelling effort.

The obtained class distribution is strongly skewed (see Table~\ref{tab:class_statistics}). This presents a fundamental challenge not only for the identification of rare morphologies but also for constructing balanced training datasets, highlighting the need for open-set approaches.

Due to the classification objective used during fine-tuning, the learned representations are not optimised to represent morphologies beyond those present in the labelled set. Adapting the model for OSR, e.g. by using a tailored loss function \citep{2023arXiv231215571S}, could improve the completeness and purity of these clusters. Nonetheless, the identification of the isolated lobes shows that this approach can recognise unexpected morphologies.

This approach is different from anomaly detection, where single sources are identified as uncommon \citep[e.g.][]{2025AJ....169..121L}. Those sources can only be used as an entire anomalous class to extend the classification scheme. In contrast, the cluster-based identification developed in this work permits the adoption of newly identified morphologies into the classification scheme (e.g., winged or FR-ambiguous classes) by providing subsets of sources with these morphologies.

The approach itself is simple and can be adopted to leverage additional pre-trained models. Potentially, multiple pre-trained models could be used for a more robust identification of rare morphologies within surveys.

In future work, the classification scheme could be extended to include additional classes and to larger source sizes (> \SI{60}{''}). The classifier ensemble could then be fine-tuned with the resulting labelled dataset, ensuring a large training set. The fine-tuning strategy could include the open-set recognition learning task. With that model, the entire LoTSS-DR3, as well as other surveys, such as EMU \citep{2025PASA...42...71H}, could be analysed.

We have shown that representation learning can be used to identify new morphologies in evolving datasets, such as those from new survey data. We demonstrate the potential of this approach using LoTSS-DR3 data, and expect it to be highly valuable for forthcoming surveys with the SKA and other telescopes.

\section{Data availability}

The catalogue of source classifications is only available in electronic form at the CDS via anonymous ftp to \url{cdsarc.u-strasbg.fr} (130.79.128.5) or via \url{http://cdsweb.u-strasbg.fr/cgi-bin/qcat?J/A+A/}.

\begin{acknowledgements}

NBP acknowledges funding by the Hamburg Virtual Initiative for Science \& Technology in AI (Hamburg VISTA), and by the Deutsche Forschungsgemeinschaft (DFG, German Research Foundation) – project number 460248186 (PUNCH4NFDI). MB acknowledges funding by the Deutsche Forschungsgemeinschaft (DFG) under Germany's Excellence Strategy -- EXC 2121 ``Quantum Universe" --  390833306 and the DFG Research Group "Relativistic Jets". LLS acknowledges funding by the Deutsche Forschungsgemeinschaft (DFG, German Research Foundation) under Germany’s Excellence Strategy -- EXC 2121 ``Quantum Universe'' -- 390833306 and via the SciFM consortium (05D25GU4) funded by the German Federal Ministry of Research, Technology, and Space (BMFTR) in the ErUM-Data action plan.

\end{acknowledgements}

\bibliographystyle{aa}
\bibliography{references}

\end{document}